\documentclass[aip,amsmath,amssymb,reprint,groupedaddress]{revtex4-1}

\usepackage{graphicx} 
\usepackage{amsmath} 
\usepackage{amssymb}   
\usepackage{xcolor}
\usepackage{color}  

\newcommand{\ket}[1]{| #1 \rangle}

\newcommand{\braket}[2]{\langle #1 | #2 \rangle}

\usepackage{physics} 
\usepackage[normalem]{ulem}
\usepackage[T1]{fontenc}

\newcommand{\rb}{^{87}{\rm Rb}}

\definecolor{mscolor}{rgb}{0,0.5,0.5}
\definecolor{akscolor}{rgb}{0.5,0,0}
\definecolor{phcolor}{rgb}{0.5,0,0.5}
\definecolor{cycolor}{rgb}{0,0.4,0}

\begin{document}

\title{An architecture for  quantum networking of neutral atom processors}
\author{C. B. Young, A. Safari, P. Huft, J. Zhang, E. Oh, R. Chinnarasu, and M. Saffman}
\affiliation{
Department of Physics,  University of Wisconsin-Madison, 
Madison, WI, 53706
}

 \date{\today}

\begin{abstract} Development of a  network for remote entanglement of quantum processors is an outstanding challenge in quantum information science. We propose and analyze a two-species architecture for remote entanglement of neutral atom quantum computers based on integration of optically trapped atomic qubit arrays with fast optics for photon collection. One of the atomic species is used for atom-photon entanglement, and the other species provides local processing. We compare the achievable rates of remote entanglement generation for two optical approaches: free space photon collection with a lens and a near-concentric, long working distance resonant cavity. Laser cooling and trapping within the cavity removes the need for mechanical transport of atoms from a source region, which allows for a fast repetition rate. Using optimized values of the cavity finesse, remote  entanglement generation rates $> 10^3~\rm s^{-1}$ are predicted for experimentally feasible parameters. 

\end{abstract}

\maketitle


\vspace{1.cm}

\section{Introduction}
An outstanding challenge in the field of quantum information science is the ability to connect multiple quantum processors or sensors in a quantum network. This capability will enable distributed quantum computation and sensing, as well as secure long distance communication based on networked quantum repeaters, each containing an interface between stationary and flying qubits, quantum memory, and processing capability. Moreover, at any stage of the development of quantum computers, there may be a practical limitation on the achievable number of qubits in a single processing unit. Therefore, quantum networks will enable a route towards building a large scale quantum processor based on a modular architecture\cite{Monroe2014}.

\begin{figure*}[!t]
    \centering
    \vspace{.5cm}
    \includegraphics[width=.8\textwidth,trim=0cm 0cm 0cm .1cm,clip=true]{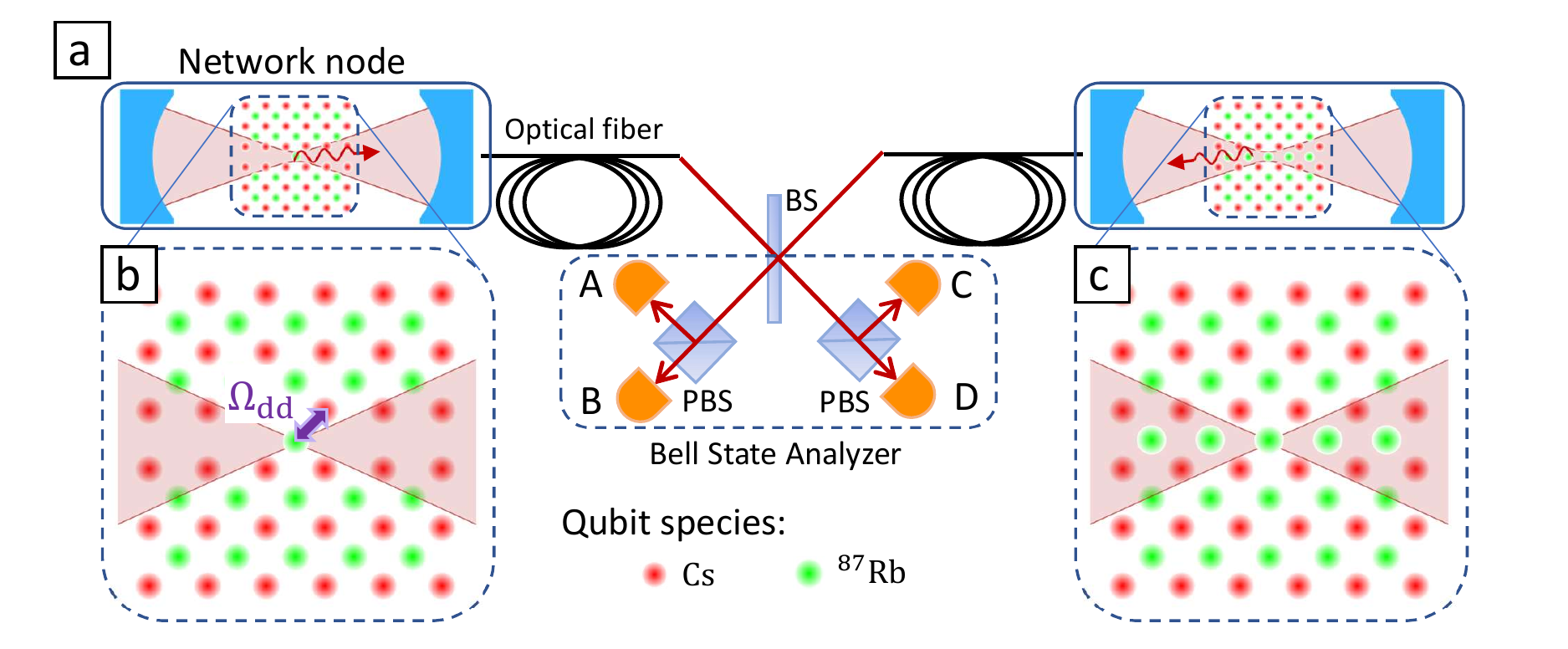}
    \caption
    {\label{fig.network} a) Illustration of two neutral atom quantum registers connected by a photonic link. Two schemes are shown for mediating the transfer of atomic excitations onto photon polarization states: b) a single atom interacting with the cavity mode, and c) multiple atoms interacting with the cavity mode. Entanglement between a register atom and communication atom(s) at each node is mediated by Rydberg interactions, indicated by $\Omega_{\text{dd}}$. Coherent scattering of photons from the single Rb atoms results in atom-photon entanglement at each node, which is projected onto atom-atom entanglement upon interference at the Bell state analyzer. The measurement requires a 50:50 beam splitter (BS), two polarizing beam splitters (PBS) and four single-photon counting modules (labeled A to D).}
\end{figure*}

The backbone of such a network is quantum entanglement distributed among the nodes. In this paper we analyze a quantum network  architecture based on interconnected functional nodes, each containing memory and processing, as well as an interface between matter  qubits and photons that mediate entanglement distribution.  A conceptual illustration of such a network based on neutral atoms is provided in Fig.\,\ref{fig.network}. Each node contains  a neutral atom quantum processor that provides memory and logic\cite{Morgado2021}. The emergent capabilities of neutral atom arrays for circuit model quantum computation have recently been established\cite{Graham2022,Bluvstein2022}. Multi-qubit logic is enabled by excitation to  Rydberg states which provide strong entangling interactions\cite{Saffman2010}. 

In order to achieve  entanglement distribution at high rates it is desirable to either collect scattered light with a high numerical aperture (NA) lens, or to strongly couple atoms and photons in a resonant cavity. Practical architectures for strong coupling of atoms to photons often involve placing the atoms close to material surfaces. This is problematic when Rydberg states are used for quantum logic since these states are strongly perturbed by electric fields arising from surface charges.  This renders nanophotonic devices for strong atom-photon coupling\cite{Asenjo-Garcia2017,Covey2019b,Menon2020,Dordevic2021} challenging to incorporate  in node architectures that utilize  Rydberg state  operations.

For this reason we  will consider approaches that have the potential to reach high rates of entanglement distribution without requiring atoms to be closer than a few mm from the nearest surface. We analyze two primary designs: free space collection of emitted photons with a high NA, yet long working distance, lens or mirror, and atom-photon coupling in a linear resonator. This leads to the type of architecture shown in  Fig.\,\ref{fig.network} where an atomic qubit array is located inside a few mm - cm scale optical resonator. In order to achieve strong coupling without requiring a very short resonator length we concentrate on a near-concentric resonator design\cite{Nguyen2017,Kawasaki2019,Huie2021}.  In a free space version the resonator is replaced by a single collection lens. Note that with the $3\,\mu\rm m$ qubit spacing demonstrated recently\cite{Graham2022}, even a large  array with $10^5$ qubits only occupies a square of size 1\,mm$^2$ and can thus be easily located inside a cm scale resonator.

A variety of approaches are possible for mediating atom-photon entanglement in each node\cite{Volz2006,Ritter2012,Uphoff2016}, including the use of Rydberg states to provide directional single photon emission\cite{Saffman2002,Saffman2005b,LLi2013,Petrosyan2018,Grankin2018}. 
The need to protect memory coherence in a multi-qubit processor while also enabling a fast qubit-photon interface is a key architectural challenge. We propose a solution based on a dual species architecture: one type of atom is used for memory and processing, and another type is used for communication and atom-photon coupling. Memory and communication qubits are coupled via interspecies Rydberg gates\cite{YZeng2017}, while the optical wavelengths used to couple to communication qubits do not perturb the memory atoms. In a two-species architecture the communication qubits can also be used for non-destructive state measurements without crosstalk to data qubits, which will enable measurement based error correction protocols\cite{Beterov2015}. This type of large, two-species atomic array was demonstrated recently\cite{Singh2022}. 
Analogous ideas have been developed for two-species trapped ion networking\cite{Inlek2017}.  Alternatively a single atomic species can be used provided that data qubits that would interact with communication light can be shelved in states that are dark with respect to transitions involved in mediating atom-photon entanglement\cite{Kwon2017}, or coherently transferred to  other states as needed for communication\cite{Covey2019,Covey2019b}.  While a detailed comparison of two-species and one-species architectures is premature at this time a few comments may be made. Laser cooling and control of two-species requires a significant overhead of additional laser systems and optics, but enables a very low level of crosstalk due to the difference in interaction wavelengths. A single species solution is likely to require less infrastructure, but may require more operations to isolate communication and processing functionalities, and suffer from higher crosstalk when implemented in the ground manifold of an alkali atom. Recent proposals for quantum networking with Yb atoms may provide more favorable single species solutions\cite{Covey2019b}.   

In the simplest version of this architecture a single communication qubit couples to the light, This can be extended to optical coupling of several communication qubits which enables generation of complex, multi-mode optical quantum states\cite{Nielsen2010}. It is also possible to combine a  sequence  of 
memory-communication qubit entanglement and communication qubit - photon entanglement steps into a single interaction where an ensemble  
of communication atoms directly mediates entanglement between a single addressed memory qubit and a photon, even though the optical wavelength does not directly interact with memory qubit transitions\cite{Wade2016}. This approach can form the basis of a multi-node network  for distributed computation\cite{Cohen2018}.

In the rest of this paper we focus on entanglement of  a single communication qubit with a photon using the protocol 
introduced in Ref.\cite{Volz2006}.
 Coherent scattering of photons by a suitably prepared $\rb$  atom leads to entanglement between the atom and a scattered photon which is collected by the aid of a lens or  optical cavity and coupled into a single-mode fiber. A Bell state measurement of the entangled photons from two nodes results in  probabilistic but heralded entanglement between the two $\rb$ atoms. Local gates between communication and data qubits at each node can then be used to entangle multi-qubit  states between nodes.

The Bell state measurement with four detectors enable remote atom-atom entanglement with a success probability of at most 50\%\cite{Calsamiglia2001}.  Although the probabilistic nature of the entanglement in this protocol does not prevent the creation of a quantum network, increasing the success probability is essential to achieve an efficient network. As we discuss below, the photon collection efficiency is often the main limiting factor for the overall success probability. Therefore, it is crucial to improve the collection efficiency to achieve a scalable quantum network. 

We compare photon collection approaches using high NA 
optics (lenses and parabolic mirrors) with optical cavities. We show that a properly designed optical cavity can improve the photon collection efficiency significantly. We consider near-concentric cavities where the large separation of the cavity mirrors allows for cooling and trapping of the atoms directly inside the cavity.

Furthermore, we analyze the case of a two-node quantum network where each node consists of a single Rb atom coupled to a near-concentric cavity. We calculate the success probability and estimate the rate of atom-atom entanglement. We show that by using a nondestructive readout of the atomic states and incorporating optical cavities, an entanglement generation rate up to a few thousand $\rm s^{-1}$ is in principle achievable.

The rest of the paper is organized as follows. In Sec. \ref{sec.lens_cavity} we compare free space and cavity geometries in terms of their photon collection efficiency. In Sec. \ref{sec.remote_entangle}  we quantify the success probability for remote entanglement of two $\rb$ atoms in a cavity geometry. In Sec. \ref{sec.rate} we analyze achievable repetition rates, and the rate of entanglement generation, accounting for the need to periodically load and recool atoms. We conclude in Sec. \ref{sec.outlook} with an outlook for implementation. 

\begin{table*}[!t]
\caption{Demonstrations of remote entanglement of matter qubits. Additional remote entanglement protocols demonstrated between ensemble quantum memories\,\cite{Choi2010, YYu2020} and with superconducting qubits and microwave fields\,\cite{Narla2016} are not included in the table.  }
\label{tab.CQED2}
\centering
\begin{tabular}{|c| l| l| c| c| c| }
\hline
Year & Research group & Description & qubit & rate $(\rm s^{-1})$& fidelity \\
\hline
2007 & Monroe \cite{Moehring2007}& remote ion-ion entanglement & $^{171}$Yb$^+$  &$0.0020~$ &0.63\\
2008 & Monroe \cite{Matsukevich2008}& remote ion-ion entanglement & $^{171}$Yb$^+$  &$0.026~$ &0.81\\
2012 & Rempe \cite{Ritter2012}& remote quantum state transfer and entanglement & $\rb$&  100& 0.85 \\
2012 & Weinfurter \cite{Hofmann2012}& atom-atom entanglement & $\rb$  & $0.0094 $ &0.81\\
2013 & Blatt \cite{Slodicka2013}& ion-ion entanglement by photon detection & $^{138}$Ba$^+$ &0.23 &0.64\\
2013 & Hanson \cite{Bernien2013}&  remote spin-spin entanglement & NV center &0.0017 & $ 0.68$ \\
2015 & Monroe \cite{Hucul2015}& remote  two ion-ion ``modular" entanglement & $^{171}$Yb$^+$  &$4.5 $ &0.78\\
2016 & Imamo\u{g}lu \cite{Delteil2016}& remote  entanglement hole spins & hole q. dots  & 2300. &0.55\\
2017 & Weinfurter \cite{Rosenfeld2017}& atom-atom entanglement, loophole free Bell test & $\rb$  & $\sim 0.03~$ &$\sim 0.85$\\
2017 & Atat\"ure \cite{Stockill2017}& remote spin-spin  entanglement & InGaAs q. dot   & $7300.$ &0.62\\
2020 & Lucas \cite{Stephenson2020}& ion-ion entanglement & $^{88}$Sr$^+$  &$182$ & 0.94\\
2021 & Hanson \cite{Pompili2021}&  three-node network & NV center &9 & $ 0.81$ \\
\hline
\end{tabular}
\end{table*} 

\section{Photon collection efficiency - Cavity vs. lens}
\label{sec.lens_cavity}

\begin{figure}[!t]
    \centering
    \vspace{.5cm}
    \includegraphics[width=.5\textwidth]{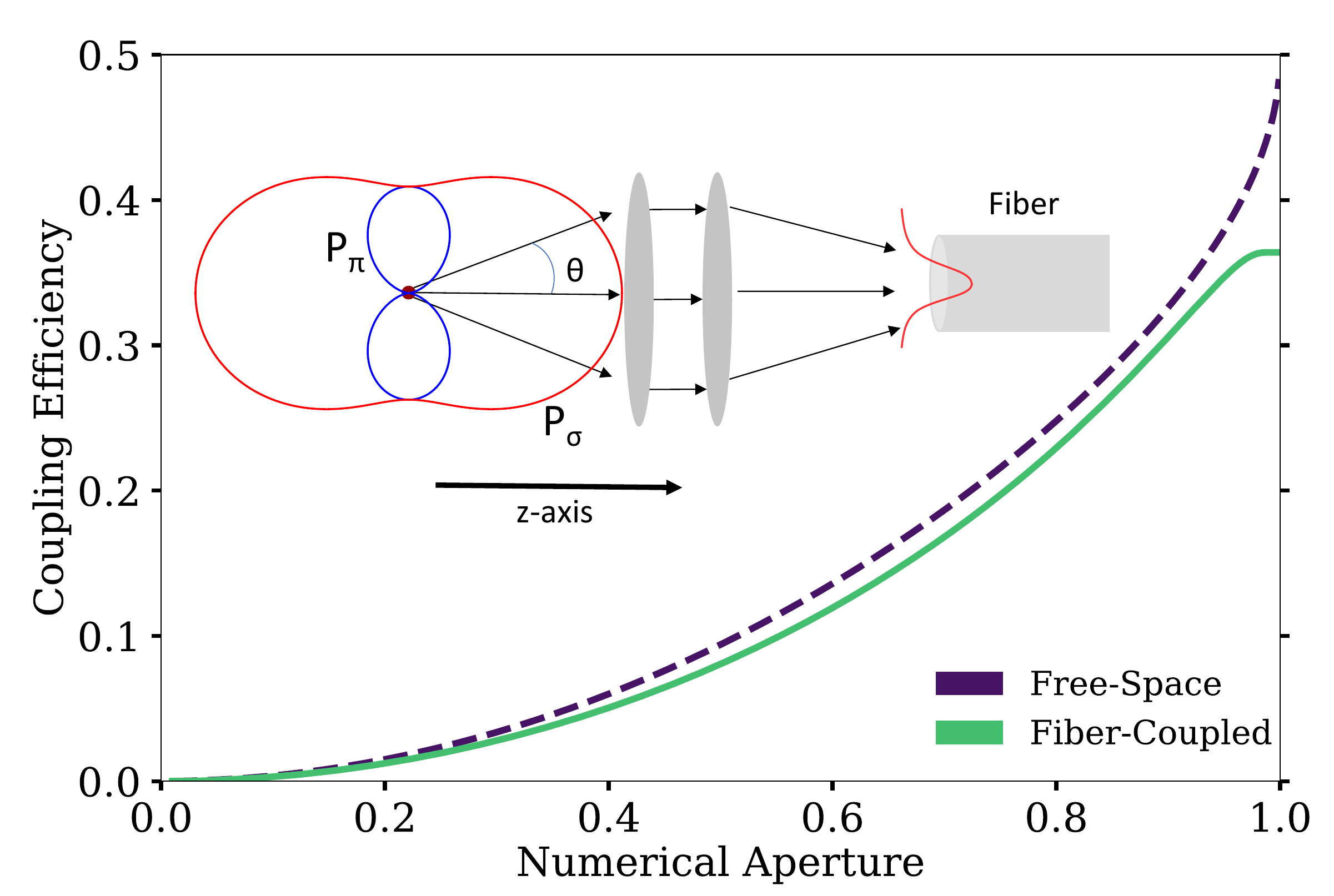}
    \caption
    {\label{fig.NA_coupling}Photon collection efficiency of a lens as a function of NA. The free-space collection efficiency (dashed line), shows the fraction of $\sigma_{\pm}$ photons collected by the lens. The fiber-coupling efficiency  from Eq.\,\eqref{eq.etalens} (solid line), is the fraction of $\sigma_{\pm}$ photons that are  coupled into a single-mode fiber. The fiber-coupled efficiency is lower due to the finite mode-overlap of the $\sigma_{\pm}$ photons and the fiber-mode at high NA. Emitted $\pi$ polarized photons do not couple into the fiber.}
\end{figure}

Efficient atom-photon coupling can be achieved with lenses or mirrors that have high NA or in resonant optical cavities. Lenses with NA up to 0.92 have been used in cold atom experiments, albeit at very short working distance\cite{Robens2017b}. This implies that the atoms are trapped very close to a surface which is problematic when using Rydberg states that have high sensitivity to electric fields. Moreover, in experiments where the polarization of the photons matters large NA optics are not always beneficial as they can lead to polarization mixing. For example, in the entanglement scheme in a polarization basis, described below, collecting the unwanted polarization  reduces the entanglement fidelity. Therefore, it is essential to couple the photons into single-mode fibers to reject the unwanted polarization. This also facilitates long-distance propagation as well as mode matching for Bell state measurements. However the fiber coupling efficiency decreases as the NA of the collection lens increases. 

In this section, we calculate the collection efficiency for circularly polarized photons with a high NA lens, as well as an optical cavity. Circularly polarized photons are considered in particular because of the entanglement scheme based on $\rb$ atoms that we consider in Sec. \,\ref{sec.remote_entangle}. The atoms also emit $\pi$-polarized photons. However, the collection efficiency of the $\pi$-polarized photons is essentially zero. When collecting the photons along the quantization axis, as shown in the inset of Fig.\,\ref{fig.NA_coupling}, due to the destructive interference of the opposite parts of the field around the symmetric axis, the $\pi$-polarized photons do not couple into the single-mode fiber. Moreover, when the lens is replaced by an optical cavity, the $\pi$-polarized photons do not couple to the cavity mode lying along the quantization axis. Therefore, in this section, we assume a two-level emitter with a $\sigma_+$ or $\sigma_-$ transition only.

In Fig.\,\ref{fig.NA_coupling} we plot the  the coupling efficiency per atomic excitation $\eta$, which is the fraction of atomic excitation cycles that lead to a photon collected by the lens and coupled into the optical fiber.  This is given by 
\begin{equation} \eta=\eta_{\rm col}O
\label{eq.etalens}
\end{equation} 
where $\eta_{\rm col}$ is the fraction of light collected by the  lens and $O$ is an overlap factor that determines the fraction of the collected light that is coupled into a propagating fiber mode.  The fiber coupling efficiency  is calculated by integrating the overlap between the dipole electric field $\mathbf{E}_{\rm D}$ of the atom and the Gaussian mode $\mathbf{E}_{\rm G}$ of the fiber
\begin{equation}
    O =\bigg\lvert \int_\Omega d\Omega \, \mathbf{E}_{\rm G} . \mathbf{E}_{\rm D}^*  \bigg\rvert ^2,
    \label{eq.overlap}
\end{equation}
where $\Omega$ is the solid angle of the collection lens.  
We assume that the quantization axis is along the optical axis of the lens. The details of the calculation are provided in Appendix\,\textbf{A}. As shown in Fig.\,\ref{fig.NA_coupling}, as the NA approaches unity, 50\% of the circularly polarized photons are collected by the lens; but less than 37\% of the photons can be coupled into the fiber due to a poor overlap with the fiber mode. Calculations for a parabolic mirror show a very similar behavior.

\begin{figure}
    \centering
    \vspace{.5cm} 
      \includegraphics[width=.5\textwidth]{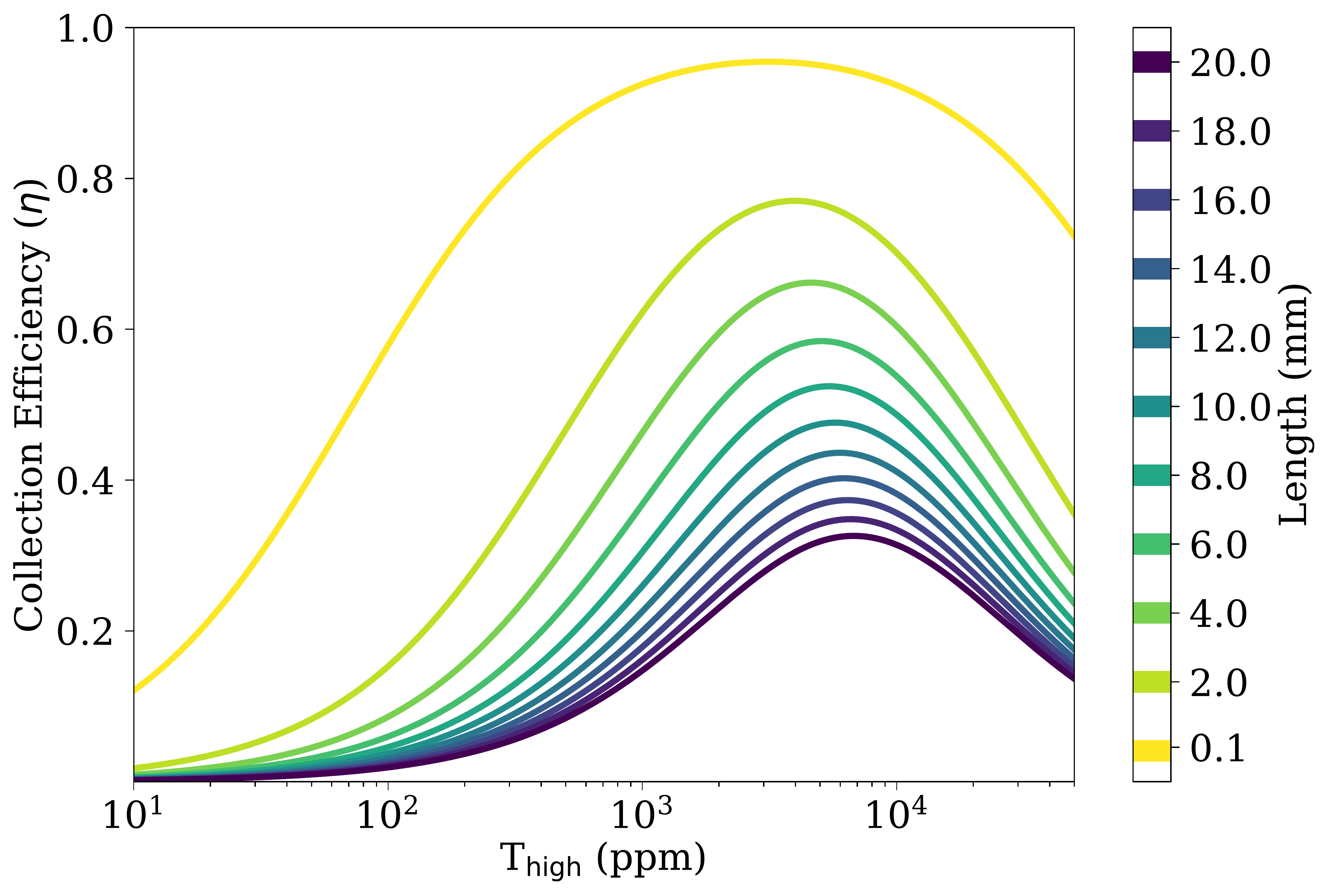}
    \caption
    {\label{fig.eta_MT_L}Achievable collection efficiency $\eta$ in an asymmetric near-concentric cavity as a function of  mirror transmission for different cavity lengths.   Parameters: $T_{\rm low}=10\,\rm ppm,$ ${\mathcal L}_{RT}= 40~$\,ppm,  $d_{\rm crit}=10~\mu\rm m$, and $\gamma/2\pi = 6.07 \times 10^6~\rm  s^{-1}$ corresponding to the $^{87}$Rb D2 line.
    }
\end{figure}

Alternatively, a resonant optical cavity can be used to enhance the atom-photon coupling by orders of magnitude, and achieve an overall collection efficiency  above 90\% in a short cavity. In atom-cavity coupling, the key parameter that captures the ratio of coherent to dissipative interactions is the cooperativity $C$ defined by~\cite{Reiserer2015}
\begin{equation}
    C=\frac{2g^2}{\kappa \gamma},
    \label{eq.coop}
\end{equation}
where $\kappa$ and $\gamma$ are the full-width at half-maximum (FWHM) of the cavity and  atomic decay rates, respectively\footnote{The equation for $C$ in\cite{Reiserer2015} appears different than our definition because those authors use $\kappa, \gamma$ defined as FWHM$/2$.}. 
The vacuum Rabi frequency is 
\begin{equation}
g= d {\mathcal E}/\hbar,
\end{equation}
where $d$ is the atomic transition matrix element, and ${\mathcal E}$ is the vacuum electric field amplitude in a mode of frequency $\omega$ in the cavity, given by
\begin{equation}
{\mathcal E}=\left(\frac{\hbar\omega}{2\epsilon_0 V} \right)^{1/2}.
\end{equation}
Here, $V= \pi w_0^2 L/4$ is the effective mode volume for a TEM$_{00}$ resonator mode of length $L$ and waist $w_0$.

\begin{figure}[t!]
    \centering 
    \includegraphics[width=.5\textwidth]{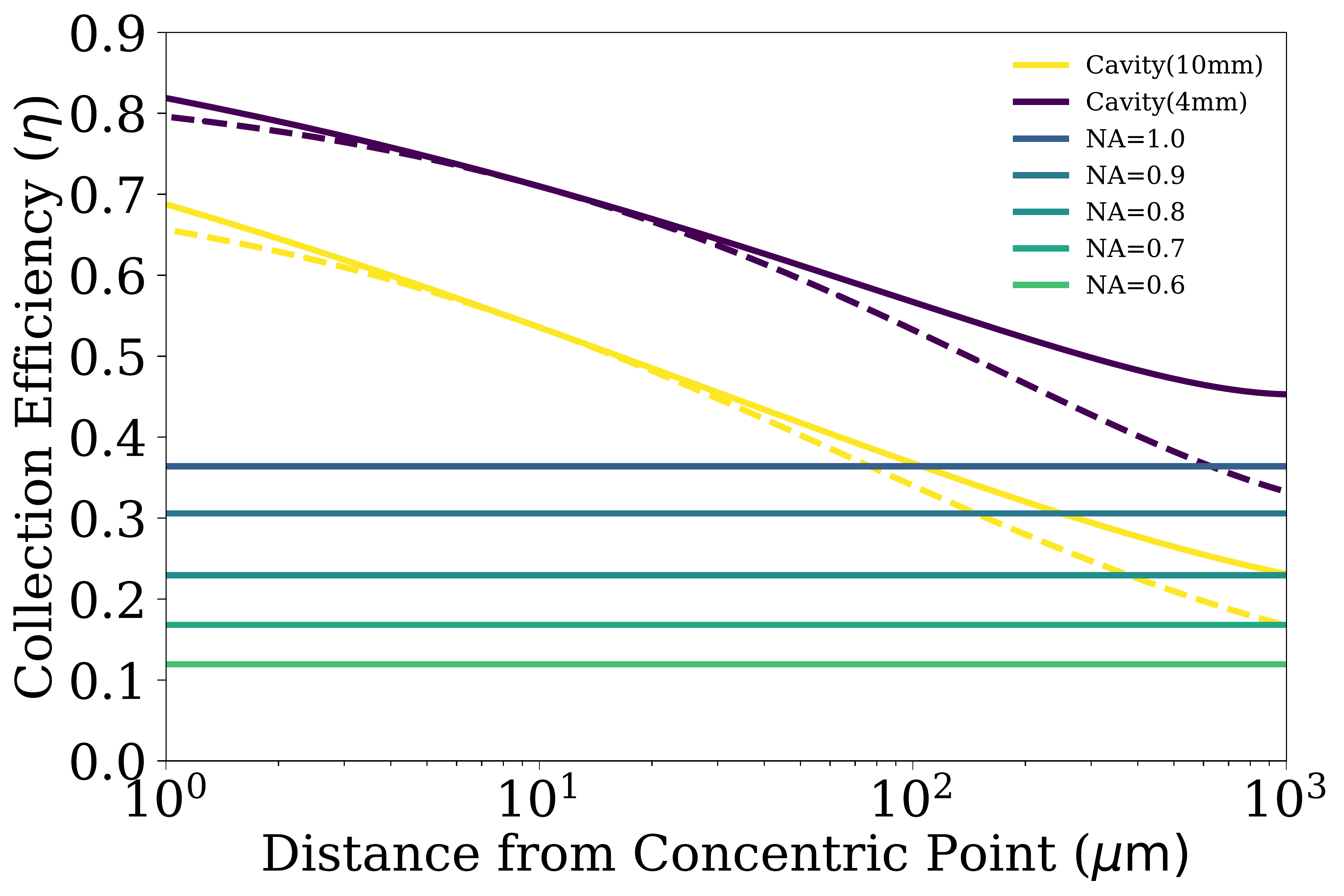}
    \caption{Photon collection efficiency comparison between resonant cavities with different lengths and high-NA lenses. The collection efficiency of the cavities is calculated as a function of the distance from the concentric point $d_{\rm crit}$.  The purple and yellow lines representing $\eta$ for the cavities are calculated using $T_{\rm low}$ = 10\,ppm, $\mathcal L_{\rm RT}$= 40\,ppm. The solid lines are calculated with $T_{\rm high}$ having a value which maximizes $\eta$ at each cavity length. The dashed lines show the value of $\eta$ with $T_{\rm high}$ optimized for a cavity with $d_{\rm crit}= 10~\mu\rm m$. We have assumed that the fiber coupling from the TEM$_{00}$ cavity mode is unity.}
    \label{fig.Lens_vs_cavity}
\end{figure}

The overall collection efficiency of the cavity is defined as the fraction of atomic excitation cycles that lead to a photon coupled into the cavity mode and transmitted through the high transmission mirror. The collection efficiency is given by
\begin{equation}
    \eta=\frac{2C}{1+2C}\frac{\kappa}{\kappa+\gamma} \frac{T_{\rm high}}{T_{\rm low} + T_{\rm high} + \mathcal L_{\rm RT}}.
    \label{eq.etaextFULL}
\end{equation}
The first factor indicates the fraction of photons coupled into the cavity mode. The second factor shows the probability that the photon leaves the cavity through one of the mirrors before being scattered again by the atom. The last fraction indicates the probability that the photon leaves the cavity through the desired mirror with high transmission of $T_{\rm high}$. Ideally, one wants the transmission of the other mirror, $T_{\rm low}$, and the round-trip loss of the cavity due to mirror absorption and scattering, $\mathcal L_{\rm RT}$, to be as small as possible. Note that in the limit of no excess cavity losses $T_{\rm low}={\mathcal L}_{\rm RT}=0$, the last term in Eq.\,\eqref{eq.etaextFULL} is unity and we recover the expression for the quantum efficiency given in \cite{GCui2005}. For all cavity geometries that we consider the mode width is several times larger than the optical wavelength, and in this limit the overlap factor of Eq.\,\eqref{eq.overlap} is essentially unity, and therefore not included. 

Equation\,\eqref{eq.etaextFULL} shows that efficient photon collection requires a relatively large cavity decay rate such that $\kappa/\gamma\gg 1 $. However, the cavity decay should occur through only one of the mirrors. Thus, a highly asymmetric cavity is desired. We also note that the cooperativity depends on the mirror parameters as well. As we will show below, by adjusting  the cavity parameters, such as the length and the mirror transmissions, the cavity can be tuned to the desired level of performance. It should be emphasized that while efficient photon collection does not depend on a particularly large $C$, other quantum operations between single atoms and photons such as state transfer or logic gates typically do require large cooperativity\cite{Daiss2021}.

The cavity energy decay rate is 
\begin{equation}
\kappa =\frac{\pi c}{L \mathcal F},
\end{equation}
where $c$ is the speed of light in vacuum, and the finesse of an optical cavity with imperfect mirrors is given by
\begin{equation}
{\mathcal F}=\frac{\pi\left[(1-T_{\rm low})(1-T_{\rm high})(1-{\mathcal L}_{\rm RT}) \right]^{1/4} }{1-\left[(1-T_{\rm low})(1-T_{\rm high})(1-{\mathcal L}_{\rm RT}) \right]^{1/2}}.
\label{eq.finesse}
\end{equation} 

To achieve strong atom-light coupling a small waist is required since the cooperativity scales as ${\mathcal F}/w_0^2$. It is often desirable to have a longer resonator without increasing the waist. This can be achieved in a near-concentric geometry for which
\begin{equation}
    w_0=w_{\rm nc}=w_{\rm c}\left(\frac{R_{\rm m}-L/2}{L/2}\right)^{1/4},
    \label{eq.nearconc}
\end{equation}
where $w_{\rm c} = \sqrt{\frac{L\lambda}{2\pi}}$ is the waist for a confocal cavity, and $R_{\rm m}$ is the mirrors radius of curvature. When $R_{\rm m}> L/2$ a stable mode with a waist much smaller than the confocal waist in a resonator of the same length  is possible.   Therefore, a near-concentric cavity offers excellent optical access for cooling and trapping atoms directly in the cavity mode while providing a small mode area necessary for strong coupling with the atoms. 

However, to achieve a small mode area, the cavity has to operate near the stability limit. The stability parameter is defined by
\begin{equation}
    s = 1 - \frac{L}{R_{\rm m}},
    \label{eq.stabilityParam}
\end{equation}
where $0 \leq s^2 \leq 1$ is required for a stable cavity. The cavity stability can be equivalently
represented by the  critical distance  $d_{\rm crit}=2R_{\rm m} - L$, i.e. the distance of the mirrors from the concentric point. Although a positive critical distance represents a stable cavity, maintaining alignment of a cavity with very small critical distance is challenging\cite{Nguyen2018}.

\begin{figure}[!t]
    \centering
    \vspace{.5cm}
    \includegraphics[width=.45\textwidth]{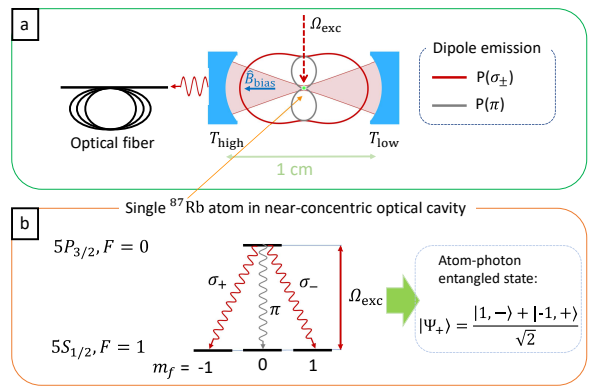}
    \caption
    {\label{fig.Rbincavity} a) Geometry for atom excitation and photon retrieval using a near-concentric optical cavity with higher mirror transmission on the fiber side. b) Relevant $\rb$  energy levels, decay and excitation scheme, and resulting atom-photon entangled state.}
\end{figure}

In Fig.\,\ref{fig.eta_MT_L} we show the photon collection efficiency of near-concentric cavities of various lengths calculated from Eq.\,\eqref{eq.etaextFULL}. We see that a collection efficiency well above 0.5 is achievable with a few-mm cavity length, and that the outcoupling mirror should have a certain transmission. With precise active alignment of the cavity mirrors near-concentric cavities with a critical distance less than a micron have been demonstrated\cite{Nguyen2018}. However, we assumed a moderate critical distance of $d_{\rm crit}=10\,\mu\rm m$. 

In Fig.\,\ref{fig.Lens_vs_cavity} we show the collection efficiency of near-concentric cavities with two different lengths as a function of the critical distance. The mirror radii of curvature for the long and short cavities are $R_{\rm m}=5$\,mm, and $R_{\rm m}=2$\,mm, respectively. As the critical distance changes, the required $T_{\rm high}$ to maximize the collection efficiency also changes. Thus, in the solid lines in Fig.\,\ref{fig.Lens_vs_cavity} we optimize $T_{\rm high}$ at each critical distance. The dashed lines in Fig.\,\ref{fig.Lens_vs_cavity} show the efficiency of the cavities with a $T_{\rm high}$ optimized for $d_{\rm crit}=10~\mu\rm m$. For comparison, the figure also includes  the collection efficiency of a lens at various numerical apertures. While the collection efficiency of a high NA lens is generally less than 30\%, it is possible to collect more than 50\% of the  scattered photons by means of an optical cavity. In addition, when multiple decay channels are accessible to the excited atom an optical cavity can enhance the decay through the desired transitions, as we discuss below.

\section{Remote entanglement protocol  with $^{87}$Rb atoms}
\label{sec.remote_entangle} 

Many different approaches to preparing remote entanglement have been proposed and demonstrated\cite{Moehring2007b}. Table \ref{tab.CQED2} lists a few earlier demonstrations of remote entanglement between matter qubits mediated by a photonic channel. Achieving high rates and high fidelity remains an outstanding challenge in the field.
In this section we provide a detailed analysis of the achievable rates using the atom-photon entanglement protocol shown in Fig.\,\ref{fig.Rbincavity}. 
A $^{87}$Rb  atom is trapped inside an optical resonator
and prepared in the level $5s_{1/2}, f=1$ by optical pumping. 
An excitation pulse of Rabi frequency $\Omega_{\rm exc}$ transfers the atom to $5p_{3/2}\ket{f=0,m_f=0}$ from which the atom decays back to the $5s_{1/2}$, $f=1$ level.  Decay into $\ket{f=1,m_f=0}$ results in emission of a $\pi$-polarized photon that does not couple to the fiber through the optical cavity. Decay into $m_f=\pm1$ is accompanied by emission of a $\sigma_\mp$ photon which can couple to the cavity. Thus the observed coherent superposition is the  entangled state 
\begin{equation}
\ket{\Psi_+}=\frac{\ket{1,\sigma_-}+\ket{-1,\sigma_+}}{\sqrt{2}}.
\end{equation}
In each ket, the first element shows the state of the atom, and the second element shows the state of the photon. The relative phase of the kets follows from the Clebsch-Gordan coefficients for the transitions.

\begin{table*}[!t]
 \caption{Summary of cavity parameters optimized for $ \rb$. Examples given are those discussed in the text: a short confocal (e.g. optical fiber-based) cavity, and two lengths of longer near-concentric cavities. While the short cavity has the highest collection efficiency $\eta$, its length is prohibitive for cooling and trapping directly in the cavity as well as using Rydberg interactions between cavity and array qubits. The longer near-concentric cavities overcome these issues, while permitting a higher $\eta$ than can be achieved with free-space optics. Other parameters are $T_{\rm low}=10$\,ppm and $\mathcal L_{\rm RT}=40$\,ppm.}
    \centering
    \begin{tabular}{|c|c|c|c|}
    \hline
     & Short confocal & Medium near-concentric & Long near-concentric \\
    \hline
     $L, R_{\rm m}, w_0$ & (150, 150, 4.3)\,$\mu$m &3.99\,mm, 2\,mm, 4.98\,$\rm\mu m$ & 9.99\,mm, 5\,mm, 6.3\,$\rm \mu m$ \\
     \hline
     $T_{\rm high}, \mathcal{F}, C$ & 1540\,ppm, 3950, 12.7  & 4620\,ppm, 1346, 3.22 & 5730\,ppm, 1080, 1.67 \\
     \hline
     $(g, \kappa, \gamma)/2\pi$ & $(98, 253, 6) \rm MHz$ &$(17, 28, 6) \rm MHz $& $(8.3, 14, 6)\, \rm MHz$ \\
     \hline
     $\eta^{(\rm Rb)}$ & 0.89 & 0.66 & 0.48 \\
     \hline
    $P_{\rm aa}$ & 0.19 & 0.1 & 0.055 \\
     \hline
    \end{tabular}
    \label{tab.smallcav}
\end{table*}

Atom-photon entanglement with this choice of atomic states was demonstrated in Ref\,\cite{Volz2006} in a free-space geometry, and later extended to atom-atom entanglement\,\cite{Hofmann2012} and a loophole-free Bell test\cite{Rosenfeld2017}. In free-space, the emission rates are the same into all three magnetic sublevels. By using a resonant cavity the atomic decay rate into $m_f=\pm1$ gets enhanced, while the decay into $m_f=0$ is not affected by the cavity. Therefore, the effect of the cavity is two-fold: increasing the collection efficiency, as explained in Sec. \ref{sec.lens_cavity}, and changing the branching ratio when multiple decay channels are available. For a transition coupled to the cavity mode, the overall decay rate is enhanced by a factor of $1+2C$ due to the Purcell effect\,\cite{Purcell1946}.
Since two of the three possible transitions couple into the cavity, $\sigma_\pm$ and $\pi$ transitions occur with probabilities given by

\begin{align}
      P_+ = P_- = & \frac{1+2C}{3+4C}\equiv P_\sigma,\label{eq.angularFactor}\\
      P_\pi= & \frac{1}{3+4C}.
\end{align}
We note that the total probability is normalized $P_+ + P_- + P_\pi = 1$. In the limit that the cooperativity $C$ is zero, each transition occurs with $1/3$ probability, as expected for transitions in free-space. Including $P_\sigma$ Eq.\,\eqref{eq.etaextFULL} is modified to 
\begin{eqnarray}
     \eta^{(\rm Rb)}&=& 2P_\sigma \eta \nonumber\\
     &=&\frac{4C}{3+4C}\frac{\kappa}{\kappa+\gamma} \frac{T_{\rm high}}{T_{\rm low} + T_{\rm high} + \mathcal L_{\rm RT}}.
\end{eqnarray}

\begin{figure}
    \centering
    \includegraphics[width = 0.45\textwidth]{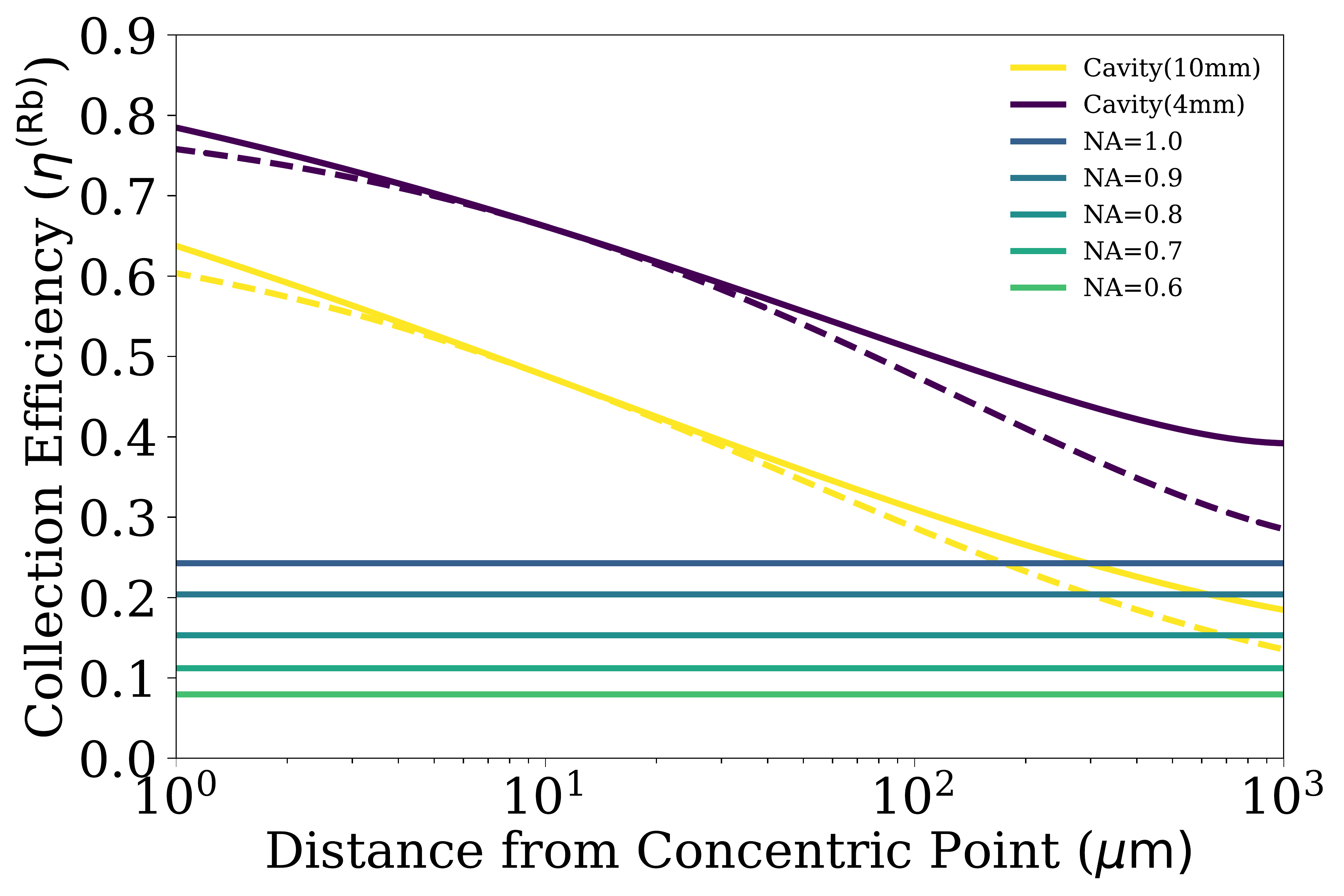}
    \caption{Photon collection efficiency comparison between resonant cavities with different lengths and high-NA lenses for the $\rb$ entanglement scheme. Parameters are identical to Fig.\,\ref{fig.Lens_vs_cavity} but now the branching ratios of the decaying atom are taken into account. The additional factor of $2P_\sigma$ makes the difference between free-space and cavity geometries  more pronounced as closer to the concentric point a higher fraction of photons emitted by the atom go into the cavity mode. For the free-space collection the branching ratio is 2/3. }
    \label{fig.Lens_vs_cavity_Rb}
\end{figure}

The dependence of $ \eta^{(\rm Rb)}$ on the cavity parameters is shown in Fig.\,\ref{fig.Lens_vs_cavity_Rb}. Compared to the case of two-level atoms shown in Fig.\,\ref{fig.Lens_vs_cavity}, the resonant cavity collection efficiencies are reduced slightly, while all the free-space collection efficiencies are reduced by a factor of 2/3. Thus, the improvement provided by a cavity is more pronounced for a real, multi-level atom.  For a 10\,mm resonant cavity with $d_{\rm crit}=10 ~\mu\rm m$, $2P_\sigma \approx 90\%$. This clearly shows that both the increased collection efficiency, and the modification of the branching ratios contribute to the better performance of resonant cavities over free-space optics.

Examples of achievable quantum efficiency are given for a short confocal and two longer near-concentric  cavity designs optimized for $^{87}$Rb in Table\,\ref{tab.smallcav}.  Short cavities with small waists provide a high quantum  efficiency. However, due to the limited optical access, laser cooling inside the cavity directly from a thermal background is impractical. Therefore, atom-transport from a cold atom source is required\,\cite{Fortier2007,Gallego2018,Hickman2020} which lowers the overall rate of photon generation. To allow for cooling inside the cavity we consider a longer 10\,mm near-concentric geometry without sacrificing the cavity mode area significantly. A medium-length 4\,mm cavity gives a significant improvement to collection efficiency over the 10\,mm cavity while still allowing for a magneto-optical trap (MOT) inside of the cavity. However, the required mirrors with small radius of curvature are not readily available.

As $d_{\rm crit}$ decreases in near-concentric cavities, the higher order spatial modes start to overlap with the fundamental mode of the cavity in frequency. Nevertheless, this is not a problem for the mm-scale cavities considered here. The frequency difference between the fundamental mode and the next transverse mode is\,\cite{Nguyen2017}
\begin{equation}
    \Delta\nu = \frac{c}{2L}\left(1-\frac{\cos^{-1}s}{\pi}\right).
    \label{eq.NCfreq}
\end{equation}
For the 10\,mm and 4\,mm cavities considered in Table\,\ref{tab.smallcav}, the frequency spacing of the modes at $d_{\rm crit}=1\,\mu$m is 95\,MHz and 377\,MHz, respectively, which are several times larger than the cavity linewidths. Therefore, even at $d_{\rm crit}$ as small as 1\,$\mu$m, the transverse modes of the cavity are still non-degenerate.

\subsection{Atom-Atom Entanglement Probability}

To generate atom-atom remote entanglement atom-photon entanglement is generated at each node, and a Bell state measurement is made at the intermediate location as shown in Fig.\,\ref{fig.network}. The probability of successful preparation of remote atom-atom entanglement on each attempt is 
\begin{eqnarray}
    P_{\rm aa}&=&\frac{1}{2} \left(\eta^{(\rm Rb)}\eta_{\rm det}\right)^2,
    \label{eq.paa}
\end{eqnarray}
where $\eta_{\rm det}$ is the quantum efficiency of the Bell state detectors, including any losses in fiber transmission. The factor of $1/2$ accounts for the success probability of the Bell state measurement.

We can make an initial estimate of $P_{\rm aa}$ using $\eta^{(\rm Rb)} = 0.48$, the quantum efficiency of the 10\,mm near-concentric cavity. The quantum efficiency of  single photon counting modules (SPCMs) at 780\,nm is $\sim$\,0.7, and the attenuation losses in the fiber can be described by $ e^{-L_{\rm f}/L_{\rm att}}$ where $L_{\rm f}$ is the fiber length in km and $L_{\rm att}= 1.091~\rm  km$ for a typical fiber at  780\,nm. We will assume a fiber length of $10~\rm m$, such that the fiber losses are negligible. Therefore, with $\eta_{\rm det} =  0.7$, we estimate that  atom-atom entanglement can be generated with a probability of $P_{\rm aa}=5.6\times 10^{-2}$.

\subsection{Remote entanglement rate}
\label{sec.rate}

\begin{table*}[!t]
\caption{Sequence of operations for a) preparing and b) verifying atom-atom entanglement. }
\label{tab.reprate}
\centering
\begin{tabular}{|l | c| c| c| }
\hline
Operation & Label &Duration & Success probability \\
\hline
\textbf{a}) \textbf{Entanglement generation:} & & &\\
\quad 1- prepare MOT, load into dipole trap &$t_{\rm load}$& 100\,ms & $> 0.99$\\
\quad 2- pump to  $5s_{1/2}\ket{1,0}$ &$t_{\rm pump}$&  $6~\mu\rm s$& $>0.99$\\
\quad 3- $\pi$-pulse to $5p_{3/2}\ket{0,0}$&$t_{\pi}$& $30\,\rm ns$&$> 0.99$ \\
\quad 4a- atomic decay, record and process  photon clicks &$t_{\rm det}$&  $1 \,{\mu\rm s}$ & $1$ \\
\quad 4b- cool atom if proper coincidence is not registered after $N_1=10$ cycles & $t_{\rm cool}$ & $100\,\mu\rm s$ & $> 0.99$ \\
\textbf{b}) \textbf{Entanglement verification:} &&&\\
\quad 5- map atomic qubit states $\ket{1,1}\rightarrow \ket{2,1}$ with 
$\mu$-wave $\pi$-pulse & $t_{\rm \mu w}$  & 50\,$\mu\rm s$& $\sim 0.99$\\
\quad 6- atomic qubit rotation with $\mu$-wave + RF $\pi/2$ pulse &$t_{\rm rot}$& 
500\,$\mu\rm s$&$ \sim0.98$\\
\quad 7- non-destructive atomic state measurement &$t_{\rm m}$& 3\,ms& $\sim 0.94$\cite{Kwon2017}\\
\quad 8- cool atom to maintain localization in trap&$t_{\rm cool}$&  100\,$\mu\rm$s& $>0.99$ \\
\hline
\end{tabular}
\end{table*}

The sequence of operations for preparing  a two-atom entangled state is presented in Table \ref{tab.reprate}. Initially an atom is loaded into each dipole trap from a MOT. This is the most time consuming step which takes approximately $t_{\rm load}$ = 100\,ms. Once  atoms have been loaded into the traps they are pumped into the $5s_{1/2}\ket{1,0}$ ground state, which  takes approximately $t_{\rm pump} = 6\, \mu\rm s$. The two atoms are then excited simultaneously with 30\,ns $\pi$-pulses to the $5p_{3/2}\ket{0,0}$ state. The SPCMs are gated for approximately 100\,ns, a few times the excitation lifetime. Proper detection of the spontaneously emitted photons from both nodes is critical for atom-atom entanglement to occur. Once the photons have been detected by the SPCMs we must determine if an appropriate photon-coincidence has occurred such that the two nodes have become entangled. This will occur with probability $P_{\rm aa}$ in each attempt following an estimated electronic processing time of $t_{\rm det} \sim 1 ~\mu\rm s$. Steps 2, 3 and 4 are repeated until a proper coincidence is registered between the SPCMs. The first four steps determine the achievable rate and how fast an entangled atom-atom pair can be generated.

To verify entanglement between the atoms we perform atom-state tomography in steps 5 through 8. First, we coherently transfer the population of $\ket{1,1}$ to $\ket{2,1}$ of the $5s_{1/2}$ ground state by a microwave $\pi$-pulse. Then, with two-photon microwave-radio frequency $\pi/2$ pulses \cite{Harber2002,Treutlein2004} we couple the states $\ket{1,-1}$ and $\ket{2,1}$ to rotate the qubit states required for  tomography. After qubit rotation, the population of $\ket{2,1}$ is measured non-destructively \cite{Kwon2017}. 

\begin{figure}
    \centering
    \includegraphics[width = 0.45\textwidth]{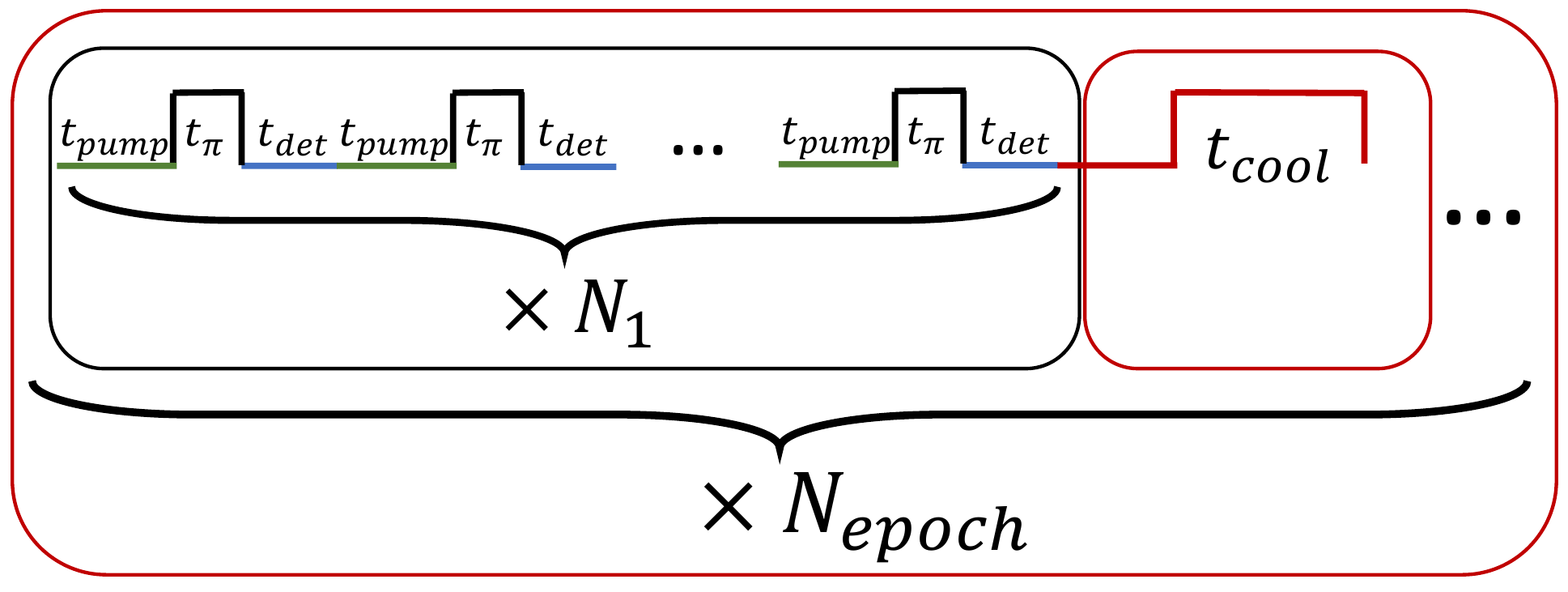}
    \caption{Addressing sequence from Eq.\, (\ref{eq.EntanglmentTime}). Referring to Table \ref{tab.reprate}, steps 2 and 3 are repeated $N_1$ times until either an entangled atom-atom pair is generated, or the atoms are cooled to continue addressing. The number of times that steps 2-4 need to be repeated on average to generate an entangled atom-atom pair is $N_{\rm epoch}$.}
    \label{fig.Addressing_Seq}
\end{figure}

Since the overall probability of establishing entanglement follows a binomial distribution, if we make $N$ attempts at entanglement generation, the expected number of successful events is $NP_{\rm aa}$, where $P_{\rm aa}$ is the atom-atom entanglement probability. Using the values in Table \ref{tab.reprate} and our estimated value of $P_{\rm aa}$, we expect a single successful event on average every $N=1/P_{\rm aa}\approx18$ attempts. The time to generate an entangled atom-atom pair once a trapped atom has been prepared is
\begin{equation}
t_{\rm aa} \simeq  N_{\rm epoch}[ N_1(t_{\rm pump}+t_{\pi}+t_{\rm det} )+t_{\rm cool}],
\label{eq.EntanglmentTime}
\end{equation}
where $N_1$ is the number of times that we address each atom between cooling cycles and $N_{\rm epoch}$ is the number of cooling cycles expected to be needed to generate atom-atom entanglement (see Fig.\,\ref{fig.Addressing_Seq}). The total number of attempts is $N=N_{\rm epoch} N_1$. We want to choose a value for $N_1$ such that the probability of loss between cooling cycles is low enough that the expected lifetime of the trapped atom is background collision limited. 
Assuming a background limited lifetime for the atom in the trap of $\sim$\,10\,s with a trap depth of $\sim\,1$\,mK, and that each pumping cycle on average will scatter less than 5 photons,
a reasonable value for $N_1$ is $\sim$10 giving a probability of loss during the period of $N_1(t_{\rm pump}+t_{\pi}+t_{\rm det})$ of $<$0.01$\%$. Assuming a cooling time of 100\,$\mu$s, this  gives an expected lifetime of $>$2\,s while continuously addressing and cooling the atom. This gives  an average  value for $N_{\rm epoch}$ of $\sim$1.8, and an expected time to generate atom-atom entanglement of
\begin{equation}
t_{\rm aa} \simeq 1.8 \times [10(t_{\rm pump} + t_{\pi} + t_{\rm det})+t_{\rm cool}] \sim 0.31 ~\rm ms,
\end{equation}
which results in an entanglement generation rate of $\sim$\,3200\,s$^{-1}$.

These estimates are based on some assumptions. The most critical are that we can address the atom without loss due to a background collision, heating  out of the trap, or depumping to $5s_{1/2},f=2$. Each time the atom is lost we would need to reload the trap resulting in $\sim$\,100\,ms of lost time, and a reduction in rate of $\sim$\,5$\%$. 
We have assumed a Doppler cooling time of 100\,$\mu$s, which has been demonstrated in similar experiments\cite{Rosenfeld2017}. The rates for these loss processes will have to be characterized and correction operations either inserted at a regular interval or performed after a fixed number of failed attempts. Leakage to $5s_{1/2},f=2$ can be rapidly corrected by shelving to $f=1$ in less than 1\,$\mu$s using light resonant with $5s_{1/2},f=2 \rightarrow 5p_{3/2}, f=1$.\\

\begin{figure}
    \centering
    \vspace{.5cm}
     \includegraphics[width=.5\textwidth]{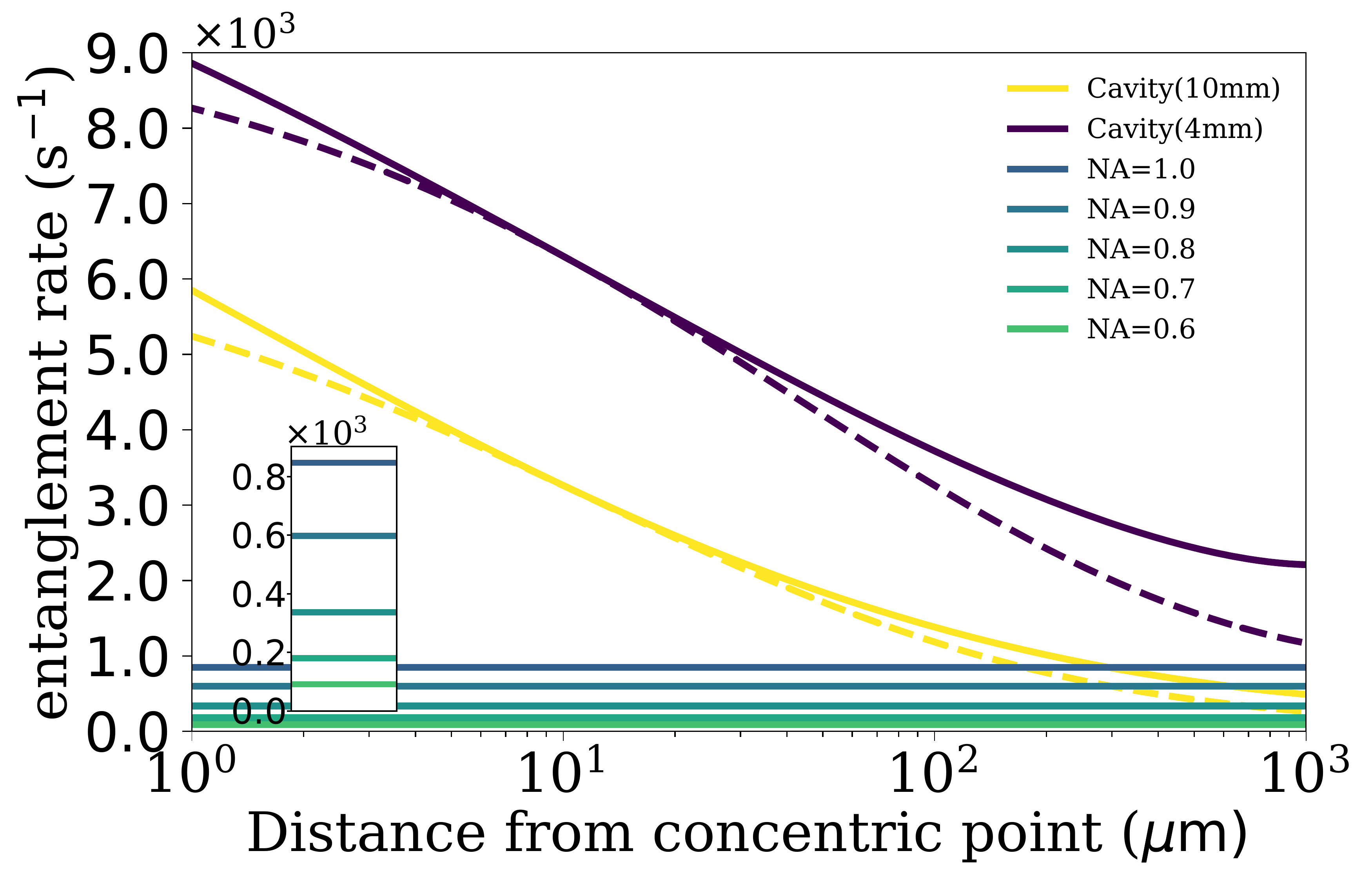}
    \caption
    {\label{fig.entanglement_rate2}Atom-atom entanglement generation rate for two cavity sizes and different high-NA lenses. The inset enlarges the rate of lenses for readability. Details are the same as Fig.\,\ref{fig.Lens_vs_cavity}.}
\end{figure}

Figure\,\ref{fig.entanglement_rate2} shows the possible rates achievable using the addressing sequence given in Eq.\,\eqref{eq.EntanglmentTime}. $t_{\rm cool}$ sets a limit on the rate assuming that cooling must happen after each entanglement generation event. The other factor which most significantly impacts the rate is the photon collection efficiency. Improving the collection efficiency will require either positioning the cavity closer to the concentric point, or using a smaller cavity. While a smaller cavity would provide a significant increase in the quantum efficiency, having a very small cavity that necessitates transporting the atoms into the cavity mode may be counter productive as the time required to do so is long and there is a large increase in complexity. Our goal is to have a cavity that is large enough that a MOT can be formed inside the cavity, and no atom-transport is required. Moving the cavity closer to the concentric point is a viable option as it has been demonstrated that a cavity in a near-concentric configuration can be maintained with $d_{\rm crit} < 1\,\mu$m using active feedback\cite{Nguyen2018}. A similar spacing for a 10\,mm cavity would decrease the mode volume by a factor of $\sim$\,7, and increase the value for $P_{\rm aa}$ by a factor of $\sim\,2$ to $1.1\times10^{-1}$, and the rate to $\sim$\,5800\,s$^{-1}$.

\subsection{Atom-atom entanglement fidelity}
\begin{table}[!t]
    \caption{Expected sources of atom-atom entanglement infidelity with their corresponding estimates. }
    \centering
    \begin{tabular}{|l|c|c|}
        \hline
         \textbf{Source of infidelity}& \textbf{Estimate($\%$)}\\
         \hline
         Temporal overlap of photons\cite{Stephenson2020} &5\\
         \hline
         Spatial overlap of photons\cite{Stephenson2020} &1\\
         \hline
         Imperfect beamsplitter and waveplates\cite{Stephenson2020} &0.2\\
         \hline
         Atom qubit rotation (for 2 atoms) &2\\
         \hline
         Atom state readout\cite{Kwon2017} (for 2 atoms) &6\\
         \hline
         Atom qubit dephasing & negligible\\
         \hline
         Detector dark count & negligible\\
         \hline
         Multi-photon scattering & negligible\\
         \hline
         Off-resonant excitation & negligible\\
         \hline
         Total & 14.2\\
         \hline
    \end{tabular}
    \label{tab:BSMInfidelitySources}
\end{table}

The fidelity of the entangled state is defined as $F=\expval{\rho}{\Psi}$, where $\ket{\Psi}$ is the maximally entangled state heralded by the photons' Bell state measurement. The density matrix $\rho$ of the entangled atom-atom state can be reconstructed through atom-state tomography. We anticipate that several factors can contribute to the loss of fidelity during the generation and verification of atom-atom entanglement. Table\,\ref{tab:BSMInfidelitySources} lists a few possible sources of infidelity with their estimated contribution. Assuming a dark count of less than 50 s$^{-1}$ we estimate that the detectors dark counts have negligible effects on the fidelity. Since the photons are coupled into single-mode fibers, a good spatial overlap at the beamsplitter is expected.

An important source of infidelity is the non-perfect temporal overlap of the photons at the beamsplitter of the Bell state analyzer. This error is $[1-\braket{\alpha_1(t)}{\alpha_2(t)}]/2$, where $\braket{\alpha_1(t)}{\alpha_2(t)}$ is the temporal overlap integral of the two photons with amplitudes $\alpha_1(t)$ and $\alpha_2(t)$. We excite the two atoms at the two nodes simultaneously with a jitter time much smaller than the decay time of the excited state. Therefore, the two photons arrive at the beamsplitter at the same time with negligible error. However, the duration of the photon wave packets depend on the exact location of the atoms with respect to the anti-node of the cavities. If an atom is displaced from the cavity anti-node, the cooperativity decreases and consequently the lifetime of the atom in the cavity increases. By using optical tweezers to hold and place the atoms in the cavities, one can control the location of the atoms with a precision that scales as $w \sqrt{k_B T_a/U}$ where $w$ is the waist of the optical tweezer, $T_a$ is the atomic temperature, and $U$ is the depth of the tweezer potential. With realistic parameters the localization can be better than  50\,nm.  Considering a more conservative error of 100\,nm for the location of the atoms, we estimate an infidelity of 5\% to due the non-perfect overlap of the photons.

Another important source of error is in non-destructive atom-state tomography. Based on  previous experiments \cite{Harber2002,Kwon2017}, we estimate that the atomic qubit can be rotated and measured with approximately 0.99 and 0.97 fidelity, respectively, at each node. Assuming a qubit coherence time of $T_2^*\sim 3~\rm ms$, which is dominated by magnetic noise and center of mass motion, and a delay of $t=60~\mu\rm s$ between photon detection and qubit measurements\cite{Stephenson2020}, we expect a loss of fidelity of order $e^{-(t/T_2^*)^2}\sim 4\times 10^{-4}$ which is negligible. Since the relative phase of the atomic $\ket{1,\pm1}$ Zeeman states is linearly sensitive to magnetic noise the local magnetic field must be well stabilized to prevent dephasing. Stabilization to better than 5 nT has been demonstrated using a feed forward approach\cite{Merkel2019} which is adequate for short photon transmission lengths.

The excitation scheme with $\pi$-polarized light prevents multi photon scattering into the cavity mode: when the atom emits a $\sigma$ polarized photon, it decays to a superposition state that is not resonantly coupled  by the $\pi$-polarized light to any excited state. Off-resonant excitation to $5p_{3/2}$ states with $f\ge 1$ is possible. However, the emitted photons from the higher excited states are not resonant to the cavity. Therefore, multi-photon scattering and off-resonant excitations do not reduce the entanglement fidelity. We also note that due to the birefringence of the optical fiber, the polarization state of the photons rotates upon propagation through the fiber. Thus, we maximize the fidelity by applying a unitary operation to the polarization state of the photons in  post processing. Also, with the assumption that the cavity subtends a small solid angle, a slight misalignment between the cavity mode and the quantization axis contributes negligibly to the entanglement infidelity. Therefore, we estimate that achieving a fidelity of approximately 0.86 is realistic with the proposed scheme. We also note that the atom state tomography, which is the dominant source of infidelity, affects only the fidelity of the verification step. Therefore, the intrinsic fidelity of atom-atom entanglement can be well above 0.9.

\section{Outlook}
\label{sec.outlook}

We have analyzed a two-species architecture  that uses one species for networking, a second species for computation, and intra-species Rydberg gates to connect the computational register with the network. A near-concentric design provides high cooperativity in a geometry compatible with in-situ atom cooling, which removes the need for atom transport, and thereby increases the  repetition rate of the entanglement generating steps. 

Assuming a short optical fiber link of 10\,m  remote entanglement rates of several hundred ${\rm s}^{-1}$ are possible using free space collection optics, and several thousand ${\rm s}^{-1}$
with resonant cavities. For longer links these rates will be reduced due to fiber attenuation. To avoid this it is necessary to convert to low loss telecom wavelengths. This can be done with appropriate atomic transitions in Rb, or other atoms\cite{Radnaev2010,Covey2019b}, or by conversion in nonlinear optical media\cite{Dreau2018}. 

In order to extend the reach of quantum networks to even longer distances multiple steps of remote entanglement are needed. The node architecture of Fig.\,\ref{fig.network} can serve as a building block for a quantum repeater based network. Incorporating the capability for writing photonic states onto communication atoms, and then performing local entangling gates, will enable deterministic Bell state measurements. In this way the factor of $1/2$ in Eq.\,\eqref{eq.paa} can be removed, which will increase the entanglement distribution rate.

\section{Acknowledgment}
This material is based upon work  supported by the U.S. Department of Energy Office of Science National Quantum Information Science Research Centers and support from  
NSF Award 2016136 for the QLCI center Hybrid Quantum Architectures and Networks.
AS acknowledges the support of the Natural Sciences and Engineering Research Council of Canada (NSERC), [funding reference number PDF - 546105 - 2020]. CBY and AS contributed equally to the manuscript. 

\bibliography{atomic,saffman_refs,rydberg,qc_refs,optics}
 
\section{Appendix}

\subsection{Fiber coupling efficiency of a high-NA lens}
\label{LensNACalc}
To calculate the coupling efficiency of the scattered photons into a single-mode fiber, we assume that the fiber supports a Gaussian mode. In other words, we assume that a Gaussian mode can be coupled into the fiber with 100\% efficiency. Thus, we find the field distribution of the radiating dipole after the collection lens, i.e. between the two lenses shown in Fig.\,\ref{fig.NA_coupling}, and calculate its overlap with a collimated Gaussian mode.

Referring to the geometry shown in the inset of Fig.\,\ref{fig.NA_coupling}, the three components of the electric field of a radiating dipole are given by
\begin{eqnarray}
    \mathbf{E}_{\sigma_{\pm}}(\theta,\phi) &=& \sqrt{\frac{3}{16\pi}} \frac{i e^{\imath (kr \pm \phi)}}{r} \left(\pm \cos{\theta} \hat{\theta} + i \hat{\phi}\right)
    \label{eq.DipoleEqSph}\\
    \mathbf{E}_{\pi}(\theta,\phi) &=& \sqrt{\frac{3}{16\pi}} \frac{i e^{\imath kr}}{r} \sin{\theta} \hat{\theta},
    \label{eq.DipoleEqSph2}
\end{eqnarray}
where $\phi$ is the azimuthal angle. After passing through the collection lens, the spherical coordinates are transformed in the following way
\begin{eqnarray}
    \hat{\theta} \rightarrow  \hat{\rho} \nonumber\\
    \hat{r} \rightarrow  \hat{z}\nonumber\\
    \hat{\phi} \rightarrow  \hat{\phi} \nonumber
\end{eqnarray}
Converting Eqs.\,\eqref{eq.DipoleEqSph} and \eqref{eq.DipoleEqSph2} to cylindrical coordinates gives the following expressions for the  field amplitudes after the collection lens
\begin{multline}
    {\mathbf{E}}_{\sigma\pm}(\rho,\phi) = \\ \sqrt{\frac{3}{16\pi}}   \frac{i e^{\pm \imath\phi}f^{1/2}}{(f^2+\rho^2)^{3/4}} \left[\frac{\pm  f}{(f^2+\rho^2)^{1/2}} \hat{\rho} + i \hat{\phi}\right],
    \label{eq.Dip1}
\end{multline}
\begin{equation}
    {\mathbf{E}}_{\pi}(\rho,\phi) =  \sqrt{\frac{3}{16\pi}}  \frac{i e^{\pm \imath\phi}f^{1/2}}{(f^2+\rho^2)^{3/4}}  \frac{\rho}{(f^2+\rho^2)^{1/2}}\hat{\rho},
    \label{eq.Dip2}
\end{equation}
where $f$ is the focal length of the lens, $\rho$ is the radial coordinate from the axis of the lens, and the factor of $f^{1/2}/(f^2+\rho^2)^{1/4}$($\sqrt{\cos{\theta}}$ in spherical coordinates before the lens) accounts for the projection of the field onto the lens plane. Note that in both coordinate systems the integral of the dipole field ${\bf E}_D={\bf E}_{\sigma\pm}$ or ${\bf E}_D={\bf E}_\pi$ over half of the full solid angle gives half of the total energy emitted by the dipole
\begin{eqnarray}
   \int_0^{\frac{\pi}{2}} d\theta\sin{\theta}  \int_{0}^{2\pi}d\phi\, \left|\mathbf{E}_{D}(\theta,\phi) \right|^2  &=& \frac{1}{2},\\
\int_{0}^{\infty} d\rho \rho\,     \int_{0}^{2\pi}d\phi\,  \left|\mathbf{E}_{D}(\rho,\phi)\right|^2  &=& \frac{1}{2}.
\end{eqnarray}

The field of a circularly polarized Gaussian laser mode with waist $w$ is given by
\begin{equation}
    \mathbf{E}_{G} = \\
    \frac{1}{\sqrt{\pi}w} e^\frac{-\rho^2}{w^2} \left[ \left(\cos \phi \pm i \sin \phi\right)\hat{\rho} - \left(\sin \phi \mp i \cos \phi\right) \hat{\phi}  \right].
    \label{eq.Gauss}
\end{equation}
We then calculate the overlap integral of the collimated dipole field, Eqs.\,\eqref{eq.Dip1} and \eqref{eq.Dip2}, with the Gaussian mode of Eq.\eqref{eq.Gauss}:
\begin{equation}
    O =\bigg\lvert
    \int_{0}^{\rho_{\rm NA}}d\rho\rho\,
    \int_{0}^{2\pi}d\phi\, \mathbf{E}_G \cdot \mathbf{E}_D^*  \bigg\rvert ^2,
    \label{eq.OInt}
\end{equation}
where $\rho_{\rm NA} = \frac{f \rm NA}{\sqrt{1-(\rm NA)^2}}$.
From this integral we see that the overlap of the $\pi$-polarized light emitted by the atom and the Gaussian mode is zero. Thus, the $\pi$-polarized photons do not couple into the single-mode fiber. We also calculate the overlap of the circularly polarized light emitted by the atom with the Gaussian mode which is shown in Fig.\,\ref{fig.NA_coupling}. As the NA of the lens changes, the beam waist $w$ in Eq.\,\eqref{eq.Gauss} has to be adjusted to maximize the overlap integral.

We note that the transverse profile of the photons collected and collimated by the lens is not Gaussian. Also, the polarization of the photons changes gradually from circular at the center ($\rho=0$) to elliptical as $\rho$ increases. Both of these effects contribute to decreasing the overlap integral in Eq.\,\eqref{eq.OInt}.

\end{document}